\documentclass[twocolumn,showpacs,showkeys]{revtex4}
\usepackage{graphicx}
\usepackage{bm}
\usepackage{color}
\usepackage{amsmath}
\usepackage{natbib}

\begin{document}

\title{Separated spin evolution quantum hydrodynamics of degenerated electrons with spin-orbit interaction and extraordinary wave spectrum}

\author{Pavel A. Andreev}
\email{andreevpa@physics.msu.ru}
\affiliation{Department of General Physics, Faculty of physics, Lomonosov Moscow State University, Moscow, Russian Federation.}
\author{Mariya Iv. Trukhanova}
\email{mar-tiv@yandex.ru}
\affiliation{Department of General Physics, Faculty of physics, Lomonosov Moscow State University, Moscow, Russian Federation.}

\date{\today}

\begin{abstract}
To consider a contribution of the spin-orbit interaction in the extraordinary wave spectrum we derive a generalization of the separate spin evolution quantum hydrodynamics. Applying corresponding nonlinear Pauli equation we include Fermi spin current contribution in the spin evolution. We find that the spectrum of extraordinary waves consists of three branches: two of them are well-known extraordinary waves and the third one is the spin-electron acoustic wave (SEAW). Earlier SEAWs have been considered in the electrostatic limit. Here we include the electromagnetic effects in their spectrum at the propagation perpendicular to the external magnetic field. We find that the SEAW spectrum considerably changes at the account of transverse part of electric field. We obtain that the separate spin evolution modifies spectrum of the well-known extraordinary waves either. A change of the extraordinary wave spectrum due to the spin-orbit interaction is obtained as well.
\end{abstract}


\maketitle



\section{\label{sec:level1} Introduction}

The separate-spin evolution quantum hydrodynamics (SSE-QHDs) has demonstrated existence of the spin-electron acoustic waves (SEAWs) \cite{Andreev PRE 15 SEAW}, \cite{Andreev AoP 15 SEAW}, \cite{Andreev 1601} and allowed to derive an equation of state for the thermal part of the spin current in the regime of degenerate electrons \cite{Andreev 1510 Spin Current}, where the distribution of electrons on the quantum states are caused by the Pauli blocking.

The SSE-QHDs was generalized to consider the spin current evolution \cite{Trukhanova PLA 15}. Equation of state for the spin current flux \cite{Trukhanova PLA 15}, \cite{Andreev IJMP B 15} was also derived via SSE-QHD \cite{Andreev 1510 Spin Current}. The Coulomb exchange interaction is included in SSE-QHD in Ref. \cite{Andreev PoP 16 soliton}.

In this paper, we continue the development and generalization of SSE-QHDs and include the spin-orbit interaction (described in sections 33 and 83 in Ref. \cite{Landau Vol.4}).

After construction of many-particle quantum hydrodynamics for the charged spin-1/2 particles with the spin-spin interaction accomplished in 2001 by Kuz'menkov and coauthors \cite{MaksimovTMP 2001}, an interest to other semi-relativistic effects arises in literature. First of all, it was consisted analysis of the spin-current interaction performed in 2007 \cite{Andreev RPJ 07} which contributes in the Lorentz force. However, more interesting results came from the account of the spin-orbit interaction considered in Ref. \cite{Andreev DSS 09} in 2009 and Ref. \cite{pavelproc} in 2011. Applications of the hydrodynamic model derived in \cite{pavelproc} were presented in Refs. \cite{Andreev IJMP B 12}, \cite{Trukhanova EPJD 13}. The spin-orbit interaction was also considered in Ref. \cite{Asenjo NJP 12}, along with the Darwin term. However, the Darwin term was considered in reduced form corresponding to the single particle motion only, but for system of identical particles it has different contribution (see discussion in Ref. \cite{Ivanov PTEP 15}). Complete result for the Darwin term or the Darwin interaction contribution is presented in Ref. \cite{Ivanov PTEP 15}. The Darwin interaction shows similarity to another interaction in the quantum semi-relativistic plasmas. This interaction arises at the simultaneous account of the Coulomb interaction and the semi--relativistic part of kinetic energy \cite{Ivanov PTEP 15}, \cite{Ivanov IJMP B 14}. They give equal contribution in the linear approximation \cite{Ivanov PTEP 15}. Spin-1/2 plasmas interacting with neutrino and appearance of an electromagnetic instability in this regime is also under consideration \cite{Bhatta arxiv 16}.

The semi-relativistic part of the quantum Bohm potential is derived in Refs. \cite{Ivanov PTEP 15}, \cite{Ivanov IJMP B 14} along with the current-current interaction giving contribution in the Lorentz force, classic semi-relativistic forces, and obtained for the first time quantum semi-relativistic forces \cite{Ivanov IJMP B 14}. Some works (see for instance Refs. \cite{Asenjo NJP 12}, \cite{Hussain PP 14 spin bernst}) are dedicated to the kinetic model of semi-relativistic effects in plasmas.
The "fluidization" of Dirac equation, via the construction of observables from bilinear covariants, was first done by Takabayasi
\cite{Takabayasi Dirac eq} and nicely reviewed in \cite{Asenjo PP 11}.
A quantum relativistic Vlasov equation was derived from the single particle Klein-Fock-Gordon \cite{Mendonca PP 11} equation and the single particle Dirac \cite{Zhu PPCF 12} equation. It was applied to the calculation of the spectrum of collective excitations (the Langmuir and electromagnetic waves) in quantum-relativistic unmagnetized plasmas. Some relativistic effects in quantum plasmas were reviewed in Ref. \cite{Uzdensky RPP 14}.

The set of QHD equations can be presented as the non-linear Schrodinger equation (NLSE) \cite{MaksimovTMP 1999}. It can be done for the spinless particles \cite{MaksimovTMP 1999} and spinning particles \cite{MaksimovTMP 2001}. In the last case we have the non-linear Pauli equation \cite{MaksimovTMP 2001}. NLSEs for the three--dimensional and two--dimensional degenerate electron gases with the Coulomb exchange interactions are derived in Ref. \cite{Andreev AoP 14}. The non-linear Pauli equation for the electron gas with the spin-orbit interaction was presented in \cite{Andreev IJMP B 12}. The set of two non-linear Pauli equations for the spin-1/2 electron-positron plasmas with the annihilation interaction was derived in \cite{Andreev PP 15 Positrons}. NLSEs, or more exactly, non-linear Klein-Fock-Gordon and non-linear Feynman--GellMaan equations was presented in Ref. \cite{Mahajan IJTP 14}. A discussion of non-linear Pauli equation can also be found in Ref. \cite{Mahajan IJTP 14}.

SEAWs are the longitudinal waves in spin-1/2 plasmas along with the Langmuir and Trivelpiece--Gould waves.
The SEAWs can be described if we consider the spin-up and spin-down electrons as two different fluids.
Moreover, it is necessary to have different equilibrium concentrations of the spin-up and spin-down electrons. Hence,
the contributions of their pressures are also different. All of these differences are caused by the partial spin polarization of electrons.
The SEAW was predicted at the wave propagation parallel and perpendicular to the external magnetic field \cite{Andreev PRE 15 SEAW}. In this regime, it has one branch of the dispersion dependence. However, there are two branches of the bulk SEAWs at the oblique propagation \cite{Andreev AoP 15 SEAW}. Influence of the spin polarization on traditional electrostatic waves is considered in Ref. \cite{Margulis JETP 87}.

The lower branch of SEAWs has zero frequency at $k=0$. It becomes the single branch at the propagation parallel to the external magnetic field $\theta=0$.
The upper branch is located above the Trivelpiece--Gould wave spectrum. Its frequency tends to $\Omega_{e}$ at $k\rightarrow0$, where $\Omega_{e}=eB_{0}/mc$ is the cyclotron frequency of electrons. It becomes the single branch of the SEAWs at the perpendicular propagation $\theta=\pi/2$ \cite{Andreev AoP 15 SEAW}.
A small Landau damping of SEAWs propagating parallel to the external magnetic field is demonstrated in Ref. \cite{Andreev 1409}, where the SSE quantum kinetics is derived.
Properties of the bulk SEAWs in the electron-positron plasmas are similar to the SEAWs in electron-ion plasmas, but the frequencies are modified due to equality of masses of electrons and positrons \cite{Andreev 1601}. The SEAWs in the electron-positron-ion plasmas, where the number of electrons and positrons are different, are also described in the regime of bulk waves. At the parallel propagation, system shows the existence of three longitudinal waves with linear spectrum: the SEAW, the positron-acoustic wave, and the spin-electron-positron acoustic wave. At the oblique propagation each of these waves has a sibling with a dispersion curve located above the Trivelpiece--Gould wave spectrum \cite{Andreev 1601}. In this paper we shift our attention on role of the SSE in the longitudinally-transverse waves such as extraordinary waves since waves with transverse field are important for different applications (see for instance \cite{Giannini CR 11}).

Collective behavior of quantum-relativistic plasmas
interacting with an intense circularly polarized electromagnetic wave is considered in
Ref. \cite{Mahajan PP 16}. It is pointed out that the spin up-down degeneracy is removed by the electromagnetic field. A modified dispersion relation for ordinary electromagnetic wave is found (see formula 75 of Ref. \cite{Mahajan PP 16}). Effects of the spin separation modify the contribution of medium in the dispersion relation. Instead of the traditional plasma frequency square they found a general term containing the SSE effects.

The nonlinear evolution of the bulk SEAWs propagating parallel to the external field leads to the soliton formation \cite{Andreev PoP 16 soliton}. This soliton is described at the application of a generalized SSE-QHD containing the Coulomb exchange interaction \cite{Andreev PoP 16 soliton}.
Surface SEAWs are described for the half-spaced plasma-like mediums \cite{Andreev 1512}. In this regime, it arises as a longitudinal wave with a linear spectrum. At the relatively small spin polarizations, the surface SEAW can linearly interact with the plasmon branch. This interaction leads to the generation of SEAWs.
The SEAWs in two dimensional structures are described in Ref. \cite{Andreev EPL 16}. They appears to be similar to the spin-plasmasons described in Refs. \cite{Ryan PRB 91}, \cite{Perez PRB 09}, \cite{Agarwal PRB 14}.

The high-temperature superconductivity is a challenging problem of the present day physics. However, the SEAWs give a contribution in this problem solution. Introducing the quanta of the SEAWs called spelnons, authors of \cite{Andreev HTSC 15} consider the electron-spelnon interaction. It is demonstrated that this interaction leads to the Cooper pair formation. Formation of the Cooper pairs leads to the phase transition to the superconductive state. The transition temperature appears in the range $T\in(0, 300)$ K depending on the concentration of electrons and their spin polarization.

The existence of the quantum part of the spin current or, in other words, the quantum Bohm potential contribution in the spin evolution equation is demonstrated by Takabayasi in Ref. \cite{Takabayasi PTP 55}. Contribution of this effect in the wave properties of quantum plasmas is considered in Ref. \cite{Trukhanova PrETP 13}, where it is applied for the spin-plasma waves.

Hydrodynamic description of spin-up and spin-down electrons was addressed in 2004 in Ref. \cite{Harabadze RPJ 04}. However, the difference of pressures for the spin-up and spin-down electrons was not considered there. Effects caused by noncoservation of electron number in each subspecies due to the spin-spin interaction was not included as well. Similar model was suggested in Ref. \cite{Brodin PRL 10 SPF} and it repeats same limitations. Complete model was later presented in Ref. \cite{Andreev PRE 15 SEAW}. Limitations of earlier work \cite{Brodin PRL 10 SPF} was also discussed in comment \cite{Andreev 1410 Comment}.

This paper is organized as follows. In Sec. II the basic model is described. It contain the non-linear Pauli equation with the spinor Fermi pressure contribution and the spin-orbit interaction. In Sec. III the general form of SSE-QHD equations with the spin-orbit interaction is derived from the NLPE. In Sec. IV we present the closed set of SSE-QHD equations arising after introduction of the velocity field. In Sec. V we study the contribution of the SSE and the spin-orbit interaction in the spectrum of extraordinary waves propagating perpendicular to the external field.
In Sec. VI a brief summary of obtained results is presented.

\section{\label{sec:level1} Model: NLPE with spin-orbit interaction}

The spin-1/2 QHD of plasmas can be directly derived from the many-particle Pauli equation \cite{MaksimovTMP 2001}, \cite{MaksimovTMP 2001 b}. This method allows to include the spin-orbit interaction \cite{pavelproc}, \cite{Andreev IJMP B 12} and other relativistic effects \cite{Andreev RPJ 07}, \cite{Ivanov PTEP 15}, \cite{Ivanov IJMP B 14}, \cite{Andreev PP 15 Positrons}. Next, the QHD equations can be represented in the form of the non-linear Schrodinger equation or the non-linear Pauli equation \cite{MaksimovTMP 2001}, \cite{Andreev IJMP B 12}, \cite{MaksimovTMP 1999}, \cite{Andreev AoP 14}, \cite{Andreev PP 15 Positrons}.

We do not use this method in our paper. We use the NLPE derived in \cite{Andreev IJMP B 12} for the spin-1/2 quantum plasmas with the spin-orbit interaction. Following Ref. \cite{Andreev 1510 Spin Current} we generalize it to include the spinor pressure. We apply the generalized NLPE to derive the SSE-QHD with the spin-orbit interaction and the Fermi spin current.

The NLPE equation arises as follows
$$\imath\hbar\partial_{t}\Phi(\textbf{r},t)=\biggl(\frac{1}{2m}\widehat{\textbf{D}}^{2}+q\varphi$$
\begin{equation}\label{susdSOI Pauli eq}  +\widehat{\pi} -\gamma\widehat{\mbox{\boldmath $\sigma$}}\textbf{B}-\frac{\gamma}{mc}(\widehat{\mbox{\boldmath $\sigma$}}\cdot(\textbf{E}\times\hat{\textbf{D}}))\biggr)\Phi(\textbf{r},t),\end{equation}
where
$\textbf{D}=\textbf{p}-q\textbf{A}/c$ and $\gamma$ is the magnetic moment of particles.

In this equation, the wave function is the spinor function and we present its explicit form (see section 56 in Ref. \cite{Landau Vol.3}):
$\Phi=\left(\begin{array}{ccc}
\Phi_{u} \\
\Phi_{d} \\
\end{array}\right)$.
Each of functions $\Phi_{u}$ and $\Phi_{d}$ can be presented as $\Phi_{s}=a_{s}e^{\imath\phi_{s}}$.
The three last terms in the NLPE (\ref{susdSOI Pauli eq}) contain the Pauli matrixes $\widehat{\sigma}^{\alpha}$.
The third term on the right-hand side describes the spinor pressure contribution
$\widehat{\pi}=\left(\begin{array}{cc} \pi_{u} & 0 \\
0 & \pi_{d} \\\end{array}\right)$,
which is a diagonal second rank spinor. It can be represented in term of the Pauli matrixes $\widehat{\pi}=\pi_{u}(\hat{\textrm{I}}+\widehat{\sigma}_{z})/2+\pi_{d}(\hat{\textrm{I}}-\widehat{\sigma}_{z})/2$, where $\hat{\textrm{I}}$ is the unit second rank spinor
$\hat{\textrm{I}}=\left(
\begin{array}{cc} 1 & 0 \\
0 & 1 \\\end{array} \right)$.
Explicit form of $\pi_{s}$ is determined by the equation of state.
In this paper, we consider the degenerate electrons. Hence, $\pi_{s}$ is determined by the Fermi pressure
$\pi_{s}=(6\pi^{2}n_{s})^{\frac{2}{3}}\hbar^{2}/2m$. Here, we consider one particle in a quantum state, instead of two particles with different spin directions in each state (see section 57 in Ref. \cite{Landau Vol.5}).

We are going to derive the SSE-QHD from the NLPE (\ref{susdSOI Pauli eq}). To this end, it is useful to present the explicit form of the NLPE via the wave functions of spin-up and spin-down electrons:
$$\imath\hbar\partial_{t}\Phi_{u}=\biggl(\frac{(\frac{\hbar}{\imath}\nabla-\frac{q_{e}}{c}\textbf{A})^{2}}{2m}+q_{e}\varphi -\gamma_{e}B_{z}+\pi_{u}$$
$$-\frac{\gamma_{e}}{mc}(E_{x}D_{y}-E_{y}D_{x})\biggr)\Phi_{u}-\gamma_{e}(B_{x}-\imath B_{y})\Phi_{d}$$
\begin{equation}\label{susdSOI Pauli Expl Up}
-\frac{\gamma_{e}}{mc}\biggl((E_{y}D_{z}-E_{z}D_{y})-\imath (E_{z}D_{x}-E_{x}D_{z})\biggr)\Phi_{d}, \end{equation}
and
$$\imath\hbar\partial_{t}\Phi_{d}=\biggl(\frac{(\frac{\hbar}{\imath}\nabla-\frac{q_{e}}{c}\textbf{A})^{2}}{2m}+q_{e}\varphi +\gamma_{e}B_{z}+\pi_{d}$$
$$+\frac{\gamma_{e}}{mc}(E_{x}D_{y}-E_{y}D_{x})\biggr)\Phi_{d}-\gamma_{e}(B_{x}+\imath B_{y})\Phi_{u}$$
\begin{equation}\label{susdSOI Pauli Expl Down}
-\frac{\gamma_{e}}{mc}\biggl((E_{y}D_{z}-E_{z}D_{y})+\imath (E_{z}D_{x}-E_{x}D_{z})\biggr)\Phi_{u}. \end{equation}
These equations describe the evolution of the spin-up electrons and the spin-down electrons independently. We can represent this evolution in terms of observables. This representation leads to the hydrodynamic equations.

\section{Hydrodynamic equations: General form}

\subsection{Continuity equations}

To make the first step in the derivation of the SSE-QHDs we define the concentrations of spin-up and spin-down electrons $n_{u}=\Phi_{u}^{*}\Phi_{u}$ and $n_{d}=\Phi_{d}^{*}\Phi_{d}$ and differentiate them with respect to time. Applying equations (\ref{susdSOI Pauli Expl Up}) and (\ref{susdSOI Pauli Expl Down}) for the time derivatives of the partial wave functions $\Phi_{u}$ and $\Phi_{d}$, we derive the continuity equations
$$\partial_{t}n_{u}+\nabla\textbf{j}_{u}=\frac{\gamma_{e}}{\hbar}\varepsilon^{z\beta\gamma}S_{\beta}B_{\gamma}$$
\begin{equation}\label{susdSOI cont eq spin UP}
-\frac{2\gamma}{\hbar c}\varepsilon^{z\mu\nu}\varepsilon^{\mu\alpha\beta}E^{\alpha}[\Phi_{d}^{*}r_{u}^{\nu}D^{\beta}\Phi_{u}+c.c.]/2m, \end{equation}
and
$$\partial_{t}n_{d}+\nabla\textbf{j}_{d}=-\frac{\gamma_{e}}{\hbar}\varepsilon^{z\beta\gamma}S_{\beta}B_{\gamma}$$
\begin{equation}\label{susdSOI cont eq spin DOWN}
+\frac{2\gamma}{\hbar
c}\varepsilon^{z\mu\nu}\varepsilon^{\mu\alpha\beta}E^{\alpha}[\Phi_{u}^{*}r_{d}^{\nu}D^{\beta}\Phi_{d}+c.c.]/2m,
\end{equation} where $\textbf{r}_{u}=\{1,\imath , 1\}$,
$\textbf{r}_{d}=\{1, -\imath , -1\}$ $q_{e}=-e$ for electrons, and
$c.c.$ means complex conjugation.

The spin densities presented in the continuity equations have the following definitions:
$S_{x}=\Phi^{*}\widehat{\sigma}_{x}\Phi=\Phi_{d}^{*}\Phi_{u}+\Phi_{u}^{*}\Phi_{d}=2a_{u}a_{d}\cos\Delta \phi$, $S_{y}=\Phi^{*}\widehat{\sigma}_{y}\Phi=\imath(\Phi_{d}^{*}\Phi_{u}-\Phi_{u}^{*}\Phi_{d})=-2a_{u}a_{d}\sin\Delta \phi$, where $\Delta \phi=\phi_{u}-\phi_{d}$.

Sum of $[\Phi_{d}^{*}r_{u}^{\alpha}D^{\beta}\Phi_{u}+c.c.]/2m$ and
$[\Phi_{u}^{*}r_{d}^{\alpha}D^{\beta}\Phi_{d}+c.c.]/2m$ gives the
nonrelativistic part of the spin current tensor
$J^{\alpha\beta}_{n.r.}=(\Phi^{+}\sigma^{\alpha} D^{\beta}\Phi
+h.c.)/2m$, where $h.c.$ means Hermitian conjugation.

During the derivation of continuity equations (\ref{susdSOI cont eq spin UP}) and (\ref{susdSOI cont eq spin DOWN}), we find the explicit forms of the particle currents: $j_{s}^{\alpha}=\frac{1}{2m}(\Phi_{s}^{*}D^{\alpha}\Phi_{s}+c.c.)+(-1)^{i_{s}}\gamma_{e}\varepsilon^{z\alpha\beta}n_{s}E_{\beta}/mc$, where $s=u$ or $d$ and $i_{s}$ is a number different for spin-up and spin-down electrons: $i_{u}=2$, $i_{d}=1$. Below, we use the non-relativistic part of the particles current $j_{0s}^{\alpha}=\frac{1}{2m}(\Phi_{s}^{*}D^{\alpha}\Phi_{s}+c.c.)$ along with $j_{s}^{\alpha}$. We introduce the velocity fields $\textbf{v}_{s}$ via the particle currents $\textbf{j}_{s}\equiv n_{s}\textbf{v}_{s}$, with the following explicit form of the velocities $\textbf{v}_{s}=\frac{\hbar}{m}\nabla \phi_{s}-\frac{q_{e}}{mc}\textbf{A}+(-1)^{i_{s}}\gamma_{e}\varepsilon^{z\alpha\beta}E_{\beta}/mc$, where we apply the phase of wave function $\Phi_{s}=a_{s}e^{\imath \phi_{s}}$.

Below, we show that at the introduction of the velocity fields, the right-hand side of the continuity equations arise in terms of hydrodynamic variables.

Summing up the partial concentrations $n_{s}$, we obtain the full concentration of electrons $n_{e}=n_{u}+n_{d}$, which should satisfy the continuity equation with the zero right-hand side. However, directly summing continuity equations (\ref{susdSOI cont eq spin UP}) and (\ref{susdSOI cont eq spin DOWN}), we find a nonzero right-hand side of the continuity equation for $n_{e}$:
$$\partial_{t}n_{e}+\nabla(\textbf{j}_{u}+\textbf{j}_{d})= \frac{2\gamma}{\hbar c}E^{\alpha}\varepsilon^{z\mu\nu}\varepsilon^{\mu\alpha\beta}\times$$
\begin{equation}\label{susdSOI cont eq full with RHS} \times[(\Phi_{d}^{*}r_{u}^{\nu}D^{\beta}\Phi_{u}+c.c.)/2m -(\Phi_{u}^{*}r_{d}^{\nu}D^{\beta}\Phi_{d}+c.c.)/2m],\end{equation}
where the right-hand side is caused by the spin-orbit interaction.
Considering the right-hand side of equation (\ref{susdSOI cont eq full with RHS}), we find that it can be presented as the divergence of a vector $-\Delta \textbf{j}$. During this calculation, we have
\begin{equation}\label{susdSOI}[\Phi_{d}^{*}r_{u}^{\nu}D^{\beta}\Phi_{u}+c.c.]- [\Phi_{u}^{*}r_{d}^{\nu}D^{\beta}\Phi_{d}+c.c.]= \varepsilon^{\nu z \lambda}\hbar \partial_{\beta}S_{\lambda}.\end{equation}
Applying $\nabla\times \textbf{E}=0$, we obtain that the right-hand side arises as a divergence of a vector. This vector $\Delta j^{\beta}$ can be constructed as a part of the particle current $\Delta j^{\beta}=-\frac{2\gamma}{\hbar c}E^{\gamma}\varepsilon^{z\mu\nu}\varepsilon^{\mu\gamma\beta}\varepsilon^{\nu z \lambda}\hbar S_{\lambda}$ which can be rewritten as
\begin{equation}\label{susdSOI Delta j} \Delta j^{\beta}=\frac{\gamma}{mc}E^{\gamma} (\varepsilon^{\beta\gamma\mu} S_{\mu}-\varepsilon^{\beta\gamma z} S_{z}).\end{equation}

Consequently, the full concentration $n_{e}$ satisfy the continuity equation with the zero right-hand sides:
\begin{equation}\label{susdSOI cont eq full with zero RHS}
\partial_{t}n_{e}+\nabla\textbf{j}=0 ,\end{equation}
where $\textbf{j}=\textbf{j}_{u}+\textbf{j}_{d}+\Delta \textbf{j}$.

Difference of the partial concentrations $n_{s}$ is the z-projection of the spin density $S_{z}=n_{u}-n_{d}$. Applying continuity equations (\ref{susdSOI cont eq spin UP}) and (\ref{susdSOI cont eq spin DOWN}), we find equation for $S_{z}$
\begin{equation}\label{susdSOI eq for Sz}
\partial_{t}S_{z}+\nabla(\textbf{j}_{u}-\textbf{j}_{d})=\frac{2\gamma_{e}}{\hbar}\varepsilon^{z\beta\gamma}S_{\beta}B_{\gamma}-\frac{2\gamma}{\hbar c}\varepsilon^{z\mu\nu}\varepsilon^{\mu\alpha\beta}E^{\alpha}J^{\nu\beta} .\end{equation}
The last term in formula (\ref{susdSOI eq for Sz}) presents the z-projection of the spin torque caused by the spin-orbit interaction. The full expression on the right-hand side of equation (\ref{susdSOI eq for Sz}) is the z-projection of the full spin torque.

\subsection{Euler equations}

Application of NLPE (\ref{susdSOI Pauli eq}) to the time evolution of the momentum density of the spin-up electrons $j_{u}^{\alpha}$ gives the following Euler equations for spin-up electrons:
$$m\partial_{t}j_{u}^{\alpha}+\partial_{\beta}\Pi_{u}^{\alpha\beta}=q_{e}n_{u}E^{\alpha}+\frac{q_{e}}{c}\varepsilon^{\alpha\beta\gamma}j_{0u}^{\beta}B^{\gamma}+F_{SOu}^{\alpha}$$
\begin{equation}\label{susdSOI Euler eq Up} +\gamma_{e}n_{u}\partial^{\alpha} B_{z} +\frac{\gamma_{e}}{2}(S_{x}\partial^{\alpha} B_{x}+S_{y}\partial^{\alpha} B_{y}) +\frac{m\gamma_{e}}{\hbar}\varepsilon^{z\beta\gamma}J^{\beta\alpha}B^{\gamma},\end{equation}
where the force field of the spin-orbit interaction $F_{SOu}^{\alpha}$ has the following form:
$$F_{SOu}^{\alpha}=+\frac{\gamma}{mc}\varepsilon^{z\alpha\beta}\partial_{t}(n_{u}E^{\beta}) -\frac{2\gamma}{\hbar c}\varepsilon^{z\mu\nu}\varepsilon^{\mu\beta\gamma}E^{\beta}j^{\nu\alpha\gamma}$$
$$-\frac{q}{2mc}\frac{\gamma}{mc}E^{\beta}B^{\delta}\varepsilon^{\alpha\gamma\delta} (S_{x}\varepsilon^{x\beta\gamma}+S_{y}\varepsilon^{y\beta\gamma}+2n_{u}\varepsilon^{z\beta\gamma})$$
$$+\frac{1}{2m}\frac{\gamma}{mc}(\partial^{\alpha}E^{\beta})
\biggl[\varepsilon^{z\beta\gamma}(\Phi^{*}_{u}D^{\gamma}\Phi_{u}+c.c.)$$
\begin{equation}\label{susdSOI F SO u via j} +\varepsilon^{x\beta\gamma}(\Phi_{u}^{*}D^{\gamma}\Phi_{d}+c.c.)
+\varepsilon^{y\beta\gamma}(\Phi^{*}_{u}(-\imath)D^{\gamma}\Phi_{d}+c.c.)\biggr],\end{equation}
where
\begin{equation}\label{susdSOI def j} j^{\alpha\beta\gamma}=\frac{1}{4m^{2}} (\Phi^{+}D^{\gamma}D^{\beta}\sigma^{\alpha}\Phi+h.c.) .\end{equation}
is a part of the spin current flux. The full expression for the nonrelativistic spin current flux is
\begin{equation}\label{susdSOI def J} J^{\alpha\beta\gamma}=\frac{1}{4m^{2}} (\Phi^{+}D^{\gamma}D^{\beta}\sigma^{\alpha}\Phi +(D^{\gamma}\Phi)^{+}D^{\beta}\sigma^{\alpha}\Phi+h.c.) .\end{equation}
In formula (\ref{susdSOI F SO u via j}), we have dropped the term proportional to $\nabla\times \textbf{E}$, since it is equal to zero in the semi--relativistic approach.

The momentum current for the spin-up electrons appears during our derivation of equation (\ref{susdSOI Euler eq Up}) in the following form:
$$\Pi_{u}^{\alpha\beta}=\frac{1}{4m}(\Phi_{u}^{*}D^{\alpha}D^{\beta}\Phi_{u}+(D^{\alpha}\Phi_{u})^{*}D^{\beta}\Phi_{u}+c.c.)$$
\begin{equation}\label{susdSOI} -\frac{\gamma_{e}}{mc}\varepsilon^{\mu\gamma\beta}E^{\gamma}[\Phi_{d}^{*}r_{u}^{\mu}D^{\alpha}\Phi_{u}+c.c.]/2m +n_{u}\nabla \pi_{u}. \end{equation}

For the spin-down electrons, the Euler equation is obtained in the following form:
$$m\partial_{t}j_{d}^{\alpha}+\partial_{\beta}\Pi_{d}^{\alpha\beta}=q_{e}n_{d}E^{\alpha} +\frac{q_{e}}{c}\varepsilon^{\alpha\beta\gamma}j_{d}^{\beta}B^{\gamma}+F_{SOd}^{\alpha}$$
\begin{equation}\label{susdSOI Euler eq Down} -\gamma_{e}n_{d}\partial^{\alpha} B_{z} +\frac{\gamma_{e}}{2}(S_{x}\partial^{\alpha} B_{x}+S_{y}\partial^{\alpha} B_{y}) -\frac{m\gamma_{e}}{\hbar}\varepsilon^{z\beta\gamma}J^{\beta\alpha}B^{\gamma},\end{equation}
with the force field of the spin-orbit interaction $F_{SOd}^{\alpha}$ acting on spin-down electrons due to interactions with spin-up and spin-down electrons:
$$F_{SOd}^{\alpha}=-\frac{\gamma}{mc}\varepsilon^{z\alpha\beta}\partial_{t}(n_{d}E^{\beta})
+\frac{2\gamma}{\hbar c}\varepsilon^{z\mu\nu}\varepsilon^{\mu\beta\gamma}E^{\beta}j^{\nu\alpha\gamma}$$
$$-\frac{q}{2mc} \frac{\gamma}{mc}E^{\beta}B^{\delta}\varepsilon^{\alpha\gamma\delta} (S_{x}\varepsilon^{x\beta\gamma}+S_{y}\varepsilon^{y\beta\gamma}-2n_{d}\varepsilon^{z\beta\gamma})$$
$$+\frac{1}{2m}\frac{\gamma}{mc}(\partial^{\alpha}E^{\beta})
\biggl[\varepsilon^{z\beta\gamma}(-\Phi^{*}_{d} D^{\gamma}\Phi_{d}+c.c.)$$
\begin{equation}\label{susdSOI F SO d via j} +\varepsilon^{x\beta\gamma}(\Phi^{*}_{d}D^{\gamma}\Phi_{u}+c.c.)
+\varepsilon^{y\beta\gamma}(\Phi^{*}_{d}\imath D^{\gamma}\Phi_{u}+c.c.)\biggr].\end{equation}

The momentum current for the spin-down electrons appears in the following form:
$$\Pi_{d}^{\alpha\beta}=\frac{1}{4m}(\Phi_{d}^{*}D^{\alpha}D^{\beta}\Phi_{d}+(D^{\alpha}\Phi_{d})^{*}D^{\beta}\Phi_{d}+c.c.)$$
\begin{equation}\label{susdSOI} -\frac{\gamma_{e}}{mc}\varepsilon^{\mu\gamma\beta}E^{\gamma}[\Phi_{u}^{*}r_{d}^{\mu}D^{\alpha}\Phi_{d}+c.c.]/2m +n_{d}\nabla \pi_{d}. \end{equation}

In the SSE-QHD we need to have equations for the evolution of $x$ and $y$ projections of the spin density. If we include the contribution of the SO interaction they arise as follows:
\begin{equation}\label{susdSOI} \partial_{t}S_{j}+\Im_{j}+\partial_{\beta} J^{j\beta}
=\frac{2\gamma_{e}}{\hbar}\varepsilon^{j\beta\gamma}S^{\beta}B^{\gamma}+T_{SOj}, \end{equation}
where $j=x,y$ and
\begin{equation}\label{susdSOI SC definition} J^{\alpha\beta}=\frac{1}{2m}(\Phi^{+}\sigma^{\alpha} D^{\beta}\Phi +h.c.)+
\frac{\gamma}{2mc}\varepsilon^{\alpha\beta\gamma}n_{e}E_{\gamma}\end{equation}
is the full spin current, and
\begin{equation}\label{susdSOI} T_{SO}^{\alpha}= -\frac{\gamma}{\hbar c}\varepsilon^{\alpha\mu\nu}\varepsilon^{\mu\gamma\beta}E_{\gamma}J^{\nu\beta} \end{equation}
is the spin torque caused by the SO interaction corresponding to the earlier works on the single fluid model of electrons \cite{Andreev DSS 09}, \cite{Andreev IJMP B 12}.

The spin current is an important characteristic of medium at spintronic description  \cite{Sinova RMP 15}, \cite{An Sci Rep 12}, \cite{Sun PRB 05}. In our model the many-particle spin current naturally arises in the spin evolution equation. It also appears in the Euler equation in terms describing nonconservation of the particle number at the account of spin-spin interaction. At the account of spin-orbit interaction, the spin current arises in the force field of spin orbit interaction and in the spin torque caused by the spin-orbit interaction. Parts of the spin current also exist in the continuity equation.

Modification of the distribution of electrons in momentum space and its influence on the equation of state under influence of the strong magnetic field are described in literature (see for instance Ref. \cite{Strickland PRD 12}). In contrast with this works we focus on different occupation of quantum states by the spin-up and spin-down electrons.

\subsection{Discussion of Euler equations}

Let us compare the obtained here results with the Euler equation found for the single fluid model of electrons considered in Refs. \cite{MaksimovTMP 2001, Koide PRC 13, Koide arXiv 16, Marklund PRL07, Brodin NJP07, Shukla RMP 11}, and especially Refs. \cite{pavelproc}, \cite{Andreev IJMP B 12} concerning the spin-orbit interaction. To this end, we need to consider the sum of two Euler equations to find the Euler equation in terms of the single fluid model.

Combining Euler equations (\ref{susdSOI Euler eq Up}) and (\ref{susdSOI Euler eq Down}), we find the following equation
$$\partial_{t}(j_{u}^{\alpha}+j_{d}^{\alpha})+\partial_{\beta}(\Pi_{u}^{\alpha\beta}+\Pi_{d}^{\alpha\beta})$$
\begin{equation}\label{susdSOI}=q_{e}n_{e}E^{\alpha} +\frac{q_{e}}{c}\varepsilon^{\alpha\beta\gamma}(j_{0u}^{\beta}+j_{0d}^{\beta})B^{\gamma}+\gamma_{e}S^{\beta}\partial^{\alpha}B^{\beta} +F_{SO}^{\alpha},\end{equation}
where
$$F_{SO,part}^{\alpha}=\frac{\gamma}{mc}\varepsilon^{z\alpha\beta}\partial_{t}(S_{z}E^{\beta}) +\frac{\gamma}{mc}\partial^{\alpha}E^{\beta}\cdot\varepsilon^{\beta\gamma\delta}J^{\delta\gamma}$$
\begin{equation}\label{susdSOI F so single fluid} -\frac{q}{mc}\frac{\gamma}{mc}\varepsilon^{\alpha\gamma\delta}E^{\beta}B^{\delta}\varepsilon^{\beta\gamma\mu}S^{\mu}. \end{equation}
The third term on the right-hand side is the spin-spin interaction presented by the fourth and fifth terms in the Euler equations (\ref{susdSOI Euler eq Up}) and (\ref{susdSOI Euler eq Down}). The last terms in the Euler equations (\ref{susdSOI Euler eq Up}) and (\ref{susdSOI Euler eq Down}) describing the force field related to non-conservation of spin-up and spin-down electrons and caused by the spin-spin interaction cancel each other. The Lorentz force is a semi-relativistic effect. Therefore, we can write $\frac{q_{e}}{c}\varepsilon^{\alpha\beta\gamma}(j_{0u}^{\beta}+j_{0d}^{\beta})B^{\gamma}\approx \frac{q_{e}}{c}\varepsilon^{\alpha\beta\gamma}(j_{u}^{\beta}+j_{d}^{\beta})B^{\gamma}\approx \frac{q_{e}}{c}\varepsilon^{\alpha\beta\gamma}j^{\beta}B^{\gamma}$. Thus, we have full Lorentz force field.

Let us consider the transformation of the force field of the spin-orbit interaction. The first terms in formulae (\ref{susdSOI F SO u via j}) and (\ref{susdSOI F SO d via j}) combine in the first term in formula (\ref{susdSOI F so single fluid}), where we use $S_{z}=n_{u}-n_{d}$. The second terms cancel each other. Each of the third terms in formulae (\ref{susdSOI F SO u via j}) and (\ref{susdSOI F SO d via j}) contain brackets with three terms. Combining these brackets we have scalar product of the double spin density vector with the Levi-Civita symbol. The result is presented by the last term in formula (\ref{susdSOI F so single fluid}). Next, we combine the last terms in formulae (\ref{susdSOI F SO u via j}) and (\ref{susdSOI F SO d via j}). Combining them we find elements of the spin current tensor explicitly presented in the Appendix. In the result, we find the second term in formula (\ref{susdSOI F so single fluid}).

Equation (\ref{susdSOI F so single fluid}) is not the complete Euler equation for the single fluid model. Above we show that the full particle current consists of three terms (\ref{susdSOI cont eq full with zero RHS}). If we consider the time evolution of the full particle current $\textbf{j}=\textbf{j}_{u}+\textbf{j}_{d}+\Delta \textbf{j}$ we need to include the time evolution of vector $\Delta \textbf{j}$. Its explicit form is given by formula (\ref{susdSOI Delta j}). Differentiating formula (\ref{susdSOI Delta j}) with respect to time and combining it with Euler equation (\ref{susdSOI F so single fluid}) we obtain the time derivative of the full particle current $\partial_{t}j^{\alpha}$ on the left-hand side. On the right-hand side we have change in the force field of the spin-orbit interaction. The second term in $\partial_{t}\Delta j^{\alpha}$ cancels the first term in the force field (\ref{susdSOI F so single fluid}). Thus, we have the first term in $\partial_{t}\Delta j^{\alpha}$ replaces the first term in (\ref{susdSOI F so single fluid}) and obtain final form for the spin-orbit interaction:
$$F_{SO}^{\alpha}=\frac{\gamma}{mc}\varepsilon^{\alpha\beta\gamma}\partial_{t}(E^{\beta}S^{\gamma}) +\frac{\gamma}{mc}\partial^{\alpha}E^{\beta}\cdot\varepsilon^{\beta\gamma\delta}J^{\delta\gamma}$$
\begin{equation}\label{susdSOI F so single fluid full} -\frac{q}{mc}\frac{\gamma}{mc}\varepsilon^{\alpha\gamma\delta}E^{\beta}B^{\delta}\varepsilon^{\beta\gamma\mu}S^{\mu}. \end{equation}
Before we present comparison of our result with the previously obtained formulae in the single fluid model, we need to mention that the first term in formula (\ref{susdSOI F so single fluid full}) can be represented in the following way. If we take the time derivative we find two terms $\frac{\gamma}{mc}\varepsilon^{\alpha\beta\gamma}(\partial_{t}E^{\beta})S^{\gamma}+\frac{\gamma}{mc}\varepsilon^{\alpha\beta\gamma}E^{\beta}\partial_{t}S^{\gamma}$. The second of them can be represented at the application of the non-relativistic part of the spin evolution equation $\partial_{t}S^{\alpha} +\Im^{\alpha}+\partial_{\beta}J^{\alpha\beta} =\frac{2\gamma_{e}}{\hbar}\varepsilon^{\alpha\beta\gamma}S^{\beta}B^{\gamma}$. Thus, the second term splits on three terms. One of them is related to the Fermi spin current which was found recently. We do not expect to find its contribution in earlier works. Other terms except the last term in formula (\ref{susdSOI F so single fluid full}) can be found in Refs. \cite{pavelproc}, \cite{Andreev IJMP B 12}. It can be shown that the last term in formula (\ref{susdSOI F so single fluid full}) can be found in the single fluid model. However, it was not reported in the mentioned papers.

\section{Hydrodynamic equations with the velocity field}

After the introduction of the velocity field in the derived above equations of the SSE-QHDs with the spin-orbit interaction we find a closed set of hydrodynamic equations. The parts of the spin-currents arisen in the terms describing the spin-orbit interaction reappear in terms of the concentrations and velocity fields of the spin-up and spin-down electrons.

Final forms of the continuity equations are the following:
$$\partial_{t}n_{s}+\nabla(n_{s}\textbf{v}_{s})=(-1)^{i_{s}}\frac{\gamma_{e}}{\hbar}\varepsilon^{\alpha\beta z}S^{\alpha}B^{\beta}$$
\begin{equation}\label{susdSOI cont eq spin UP vel field}
-\frac{2\gamma}{\hbar c}\varepsilon^{z\mu\nu}\varepsilon^{\mu\alpha\beta}E^{\alpha} \biggl(\frac{1}{2}v_{s}^{\beta}S^{\nu}-(-1)^{i_{s}}\frac{\hbar}{4m}\frac{\partial^{\beta}n_{s}}{n_{s}}\varepsilon^{z\nu\delta}S^{\delta}\biggr).  \end{equation}
The last term in the continuity equation is caused by the spin-orbit interaction. At the introduction of the velocity field, we find that the right-hand side of the continuity equation represent itself via the hydrodynamic variables and we are getting closer to a closed set of equations.

Final forms of the Euler equations are the following:
$$mn_{s}(\partial_{t}+\textbf{v}_{s}\nabla)\textbf{v}_{s}+\nabla p_{s}
-\frac{\hbar^{2}}{4m}n_{s}\nabla\Biggl(\frac{\triangle n_{s}}{n_{s}}-\frac{(\nabla n_{s})^{2}}{2n_{s}^{2}}\Biggr)$$
\begin{equation}\label{susdSOI Euler eq spin UP with vel} =q_{e}n_{s}\biggl(\textbf{E}+\frac{1}{c}[\textbf{v}_{s},\textbf{B}]\biggr)+\textbf{F}_{SSs}+\tilde{\textbf{F}}_{SOs},\end{equation}
with the thermal or Fermi pressure $p_{s}$, the force field of spin-spin interaction
$$\textbf{F}_{SSs}=(-1)^{i_{s}}\gamma_{e}n_{u}\nabla B_{z} +\frac{\gamma_{e}}{2}(S_{x}\nabla B_{x}+S_{y}\nabla B_{y})$$
\begin{equation}\label{susdSOI} +(-1)^{i_{s}}\frac{m\gamma_{e}}{\hbar} \varepsilon^{\beta\gamma z}\biggl(\textbf{J}_{(M)\beta}B^{\gamma}
-\textbf{v}_{s}S^{\beta}B^{\gamma}\biggr),\end{equation}
and the force field of spin-orbit interaction acting on spin-up electrons
$$F_{SOu}^{\alpha}=+\frac{\gamma}{mc}\varepsilon^{z\alpha\beta}\partial_{t}(n_{u}E^{\beta}) -\frac{2\gamma}{\hbar c}\varepsilon^{z\mu\nu}\varepsilon^{\mu\beta\gamma}E^{\beta}j^{\nu\alpha\gamma}  $$
$$-\frac{q}{2mc}\frac{\gamma}{mc}E^{\beta}B^{\delta}\varepsilon^{\alpha\gamma\delta} (S_{x}\varepsilon^{x\beta\gamma}+S_{y}\varepsilon^{y\beta\gamma}+2n_{u}\varepsilon^{z\beta\gamma})$$
$$+\frac{1}{2}\frac{\gamma}{mc}(\partial^{\alpha}E^{\beta})
\biggl[\varepsilon^{x\beta\gamma}v_{d}^{\gamma}S^{x}
+\varepsilon^{y\beta\gamma} v_{d}^{\gamma}S^{y}+\varepsilon^{z\beta\gamma}v_{u}^{\gamma}S^{z}\biggr]$$
\begin{equation}\label{susdSOI F SO u via vel}
+\frac{\hbar}{4m}\frac{\partial^{\gamma}n_{d}}{n_{d}}\frac{\gamma}{mc}(\partial^{\alpha}E^{\beta})
\varepsilon^{\mu\beta\gamma}\varepsilon^{z\mu\delta}S^{\delta},\end{equation}
\emph{and}
the force field of spin-orbit interaction acting on spin-down electrons
$$F_{SOd}^{\alpha}=-\frac{\gamma}{mc}\varepsilon^{z\alpha\beta}\partial_{t}(n_{d}E^{\beta})
+\frac{2\gamma}{\hbar c}\varepsilon^{z\mu\nu}\varepsilon^{\mu\beta\gamma}E^{\beta}j^{\nu\alpha\gamma}$$
$$-\frac{q}{2mc} \frac{\gamma}{mc}E^{\beta}B^{\delta}\varepsilon^{\alpha\gamma\delta} (S_{x}\varepsilon^{x\beta\gamma}+S_{y}\varepsilon^{y\beta\gamma}-2n_{d}\varepsilon^{z\beta\gamma})$$
$$+\frac{1}{2}\frac{\gamma}{mc}(\partial^{\alpha}E^{\beta})
\biggl[\varepsilon^{x\beta\gamma}v_{u}^{\gamma}S^{x}
+\varepsilon^{y\beta\gamma} v_{u}^{\gamma}S^{y}+\varepsilon^{z\beta\gamma}v_{d}^{\gamma}S^{z}\biggr]$$
\begin{equation}\label{susdSOI F SO d via j} -\frac{\hbar}{4m}\frac{\partial^{\gamma}n_{u}}{n_{u}}\frac{\gamma}{mc}(\partial^{\alpha}E^{\beta})
\varepsilon^{\mu\beta\gamma}
\varepsilon^{z\mu\delta}S^{\delta}.\end{equation}
A presentation of tensor $j^{\nu\alpha\gamma}$ in terms of the velocity fields is discussed in Appendix B.

In terms of the velocity field the spin current tensor arises as
\begin{equation}\label{susdSOI Spin current x with vel} J_{j\alpha}=\frac{1}{2}(v_{u}^{\alpha}+v_{d}^{\alpha})S_{j}
-\frac{\hbar}{4m} \varepsilon^{j\beta z}
\biggl(\frac{\partial^{\alpha} n_{u}}{n_{u}}-\frac{\partial^{\alpha} n_{d}}{n_{d}}\biggr)S_{\beta}. \end{equation}
The relativistic part of the spin current tensor is hidden in the definition of velocity fields.

In the Euler equations (\ref{susdSOI Euler eq spin UP with vel}) we have used a reduced form of the spin current $\textbf{J}_{(M)x}$ and $\textbf{J}_{(M)y}$ which means $J^{x\alpha}$ and $J^{y\alpha}$ correspondingly. Here, the bold symbols means a vector related to the second index in $J^{\alpha\beta}$.

After the introduction of the velocity field, the spin evolution equations have the following form:
$$\partial_{t}S_{j}+\frac{1}{2}\nabla[S_{j}(\textbf{v}_{u}+\textbf{v}_{d})]
-\frac{\hbar}{4m}\varepsilon^{j\beta z}\nabla\Biggl(S^{\beta}\biggl(\frac{\nabla n_{u}}{n_{u}}-\frac{\nabla n_{d}}{n_{d}}\biggr)\Biggr)$$
\begin{equation}\label{susdSOI eq for Sx}  +\Im_{j}
=\frac{2\gamma_{e}}{\hbar}\varepsilon^{j\beta\gamma}S^{\beta}B^{\gamma}+T_{SOj},\end{equation}
where $j=x,y$.

The set of SSE-QHD equations is coupled with the quasi-static Maxwell equations $\nabla \cdot\textbf{B}=0$, $\nabla\times \textbf{E}=0$,
\begin{equation}\label{susdSOI div E} \nabla \cdot\textbf{E}=4\pi(en_{i}-en_{eu}-en_{ed}),\end{equation}
\begin{equation}\label{susdSOI rot B}
\nabla\times \textbf{B}=\frac{4\pi q_{e}}{c}(n_{eu}\textbf{v}_{eu}+n_{ed}\textbf{v}_{ed})+4\pi\nabla\times \textbf{M}_{e},\end{equation}
where $\textbf{M}_{e}=\{\gamma_{e}S_{ex}, \gamma_{e}S_{ey}, \gamma_{e}(n_{eu}-n_{ed})\}$ is the magnetization. We consider the motionless ions. Hence, we do not include their current in equation (\ref{susdSOI rot B}).

Since we consider the degenerate electrons we use the Fermi pressures for spin-up and spin-down electrons as equations of state:
\begin{equation}\label{susdSOI Eq State partial} p_{s}=\frac{(6\pi^{2})^{2/3}}{5}\frac{\hbar^{2}}{m}n_{s}^{5/3}.\end{equation}

Vector $\mbox{\boldmath $\Im$}$ in the spin evolution equations (\ref{susdSOI eq for Sx}) is the divergence of the thermal part of the spin current tensor $\Im^{\alpha}=\partial_{\beta}J^{\alpha\beta}_{th}$.
In accordance with the NLPE equation, as it was demonstrated in Ref. \cite{Andreev 1510 Spin Current}, we have
\begin{equation}\label{susdSOI spin current many part Vector}
\mbox{\boldmath $\Im$}=(\pi_{u}-\pi_{d})\gamma[\textbf{S}, \textbf{e}_{z}]/\hbar. \end{equation}
Due to its nature we can also call it the Fermi spin current.

\section{Linearized SSE-QHD equations}

Analysis if the wave properties requires to consider the
linearized set of SSE-QHD equations. In the linear approximation,
the SSE-QHD looks relatively simple:
\begin{equation}\label{susdSOI cont lin s} \partial_{t}\delta n_{s}+n_{0s}\nabla\delta \textbf{v}_{s}=0, \end{equation}
$$mn_{0s}\partial_{t}\delta v_{s}^{\alpha}+\partial^{\alpha}\delta p_{s}=q_{e}n_{0s}\delta E^{\alpha}
\pm\gamma_{e}n_{0s}\partial^{\alpha}\delta B_{z}$$
$$+q_{e}n_{0s}\frac{1}{c}B_{0}\varepsilon^{\alpha\beta z}\delta v_{s}^{\beta}
\mp\frac{q_{e}}{mc}\frac{\gamma}{mc}n_{0s}B_{0}\varepsilon^{\alpha\gamma z}\varepsilon^{\beta\gamma z}\delta E^{\beta}$$
\begin{equation}\label{susdSOI Euler lin u}
\mp\frac{2\gamma}{\hbar
c}\varepsilon^{z\mu\nu}\varepsilon^{\beta\gamma\mu}j_{0}^{\nu\alpha\gamma}\delta
E^{\beta} \pm\frac{\gamma}{mc}n_{0s}\varepsilon^{\alpha\beta
z}\partial_{t}\delta E^{\beta}, \end{equation}
with the upper sign
for spin-up electrons and the lower sign for spin-down electrons,
and
\begin{equation}\label{susdSOI eq for Sx lin} \partial_{t}\delta S_{j} +\delta\Im_{j}
=\frac{2\gamma_{e}}{\hbar}(\varepsilon^{j\beta z}B_{0z}\delta S_{\beta}-\varepsilon^{j\beta z}S_{0z}\delta B_{\beta})\end{equation}
for the spin evolution, where we neglect the quantum Bohm potential, $S_{0z}=n_{0u}-n_{0d}$ and $\delta\Im_{\alpha}=(\pi_{0u}-\pi_{0d})\gamma[\delta\textbf{S}, \textbf{e}_{z}]/\hbar$, with $\pi_{0s}=(6\pi^{2}n_{0s})^{\frac{2}{3}}\hbar^{2}/2m$.

The spin-orbit interaction describes the force acting on a magnetic moment moving in an external electric field. It exists even for the static electric fields. Therefore, the spin-orbit interaction affects the evolution of electrostatic waves in the spin-1/2 plasmas, since electrons possess the magnetic moments filling the electric field caused by the charges of surrounding electrons.

The fourth term on the right-hand side of each Euler equation is proportional to $\varepsilon^{z\mu\nu}\varepsilon^{\beta\gamma\mu}j_{0}^{\nu\alpha\gamma}$. The first index in $j_{0}^{\nu\alpha\gamma}$ describes the spin. In the equilibrium state, all spins are parallel to the external field. Hence, index $\nu$ is equal to $z$. Consequently, we obtain $\varepsilon^{z\mu z}\varepsilon^{\beta\gamma\mu}j_{0}^{z\alpha\gamma}$ and find that the first Levi-Civita symbol is equal to zero. Thus, this term gives zero contribution in the linear wave evolution.

The last term in Euler equations is proportional to $\varepsilon^{\alpha\gamma z}\varepsilon^{\beta\gamma z}\delta E^{\beta}=\delta E^{\alpha}-\delta^{\alpha z}\delta E^{z}$. We find the following results for different projections: $\delta E^{x}$ if $\alpha=x$ since the second term is equal to zero, similarly $\delta E^{y}$ if $\alpha=y$, $0$ if $\alpha=z$ since two terms are cancel each other.

Semirelativistic effects, such as the spin-orbit interaction,
arise in the electro-magneto-static approximation as it is shown
in formulae (\ref{susdSOI div E}) and (\ref{susdSOI rot B}). If we
want to consider the influence of the semirelativistic effects on
the propagation of the electromagnetic waves in magnetized plasmas
we need to use the full set of Maxwell equations. Here, we present
them in the linearized form $\nabla \cdot \delta\textbf{B}=0$,
$\nabla\times \delta\textbf{E}=-\partial_{t}\delta \textbf{B}/c$,
\begin{equation}\label{susdSOI div E lin} \nabla \cdot\delta\textbf{E}=-4e\pi(\delta n_{eu}+\delta n_{ed}),\end{equation}
\begin{equation}\label{susdSOI rot B lin}
\nabla\times \delta\textbf{B}=-\frac{4\pi
e}{c}(n_{0u}\delta\textbf{v}_{u}+n_{0d}\delta\textbf{v}_{d})+4\pi\gamma_{e}\nabla\times
\delta\textbf{S}_{e}.\end{equation} The application of the
linearized Maxwell equations together with the hydrodynamic
equations (\ref{susdSOI cont lin s}), (\ref{susdSOI Euler lin u}),
(\ref{susdSOI eq for Sx lin}), we study the dispersion dependence
of the extraordinary waves in the next section.

\section{Spin-orbit interaction contribution in the spectrum of longitudinal waves}

It has been recently demonstrated that longitudinal spin waves, called the spin-electron acoustic waves, can exist in the degenerate electron gas \cite{Andreev PRE 15 SEAW}. The oblique propagation of longitudinal waves reveals the existence of the second or upper SEAW \cite{Andreev AoP 15 SEAW}. The partial spin polarization is required for the existence of the SEAWs. It is also necessary to apply the SSE-QHD. The SSE-QHD is generalized in this paper. The explicit analytical contribution of the spin-orbit interaction in the spectrum of the Langmuir waves propagating perpendicular to the external magnetic field (the upper hybrid wave) was found in Ref. \cite{Ivanov PTEP 15}. It shows that the spin-orbit interaction gives the contribution in the spectrum of the longitudinal waves. This contribution arises from the following term: $F_{1}^{\alpha}=-\frac{\gamma}{mc}S_{0z}\varepsilon^{\alpha\beta z}\partial_{t}\delta E^{\beta}$. In this paper, we generalize the force field of spin-orbit interaction including the following term: $F_{2}^{\alpha}=-\frac{q_{e}}{mc}\frac{\gamma}{mc}S_{0z}B_{0}\varepsilon^{\alpha\gamma z}\varepsilon^{\beta\gamma z}\delta E^{\beta}$. They do not equal to each other. However, in the linear approximation they cancel contribution of each other in the dispersion dependence of the longitudinal waves. Therefore, we do not expect any contribution of the spin-orbit interaction in the spectra of oblique propagating longitudinal waves: Langmuir, Trivelpiece--Gould, lower and upper SEAWs. However, the spin-orbit interaction affects these waves if we do not apply restriction to consider longitudinal waves. In general case they are longitudinal--transverse waves and the spin-orbit interaction gives contribution via the transverse part.

\section{Dispersion dependance of waves propagating perpendicular to the external magnetic field}

Let us present the dispersion equation for the waves propagating perpendicular to the external magnetic field
$$ \omega^{2}-k^{2}c^{2}
-\sum_{s=u,d}\frac{\omega_{Le}^{2}}{\omega^{2}-\Omega_{e}^{2}-k^{2}U_{Fs}^{2}}\times(2\omega^{2}-\omega_{Le}^{2}$$
\begin{equation}\label{susdSOI Disp eq} -k^{2}c^{2}
-k^{2}U_{Fs}^{2}
+\alpha_{e}\Omega_{e}(2\omega_{Le}^{2}+k^{2}U_{Fs}^{2}))=0
\end{equation}
found in this paper.

The SSE affects the ordinary electromagnetic wave and spin-plasmas wave described by $\Lambda_{zz}$ via the Fermi spin current. This effect is described in Ref. \cite{Andreev 1510 Spin Current}. Therefore, we do not discuss the ordinary electromagnetic wave and spin-plasmas wave in this paper.

\begin{figure}
\includegraphics[width=8cm,angle=0]{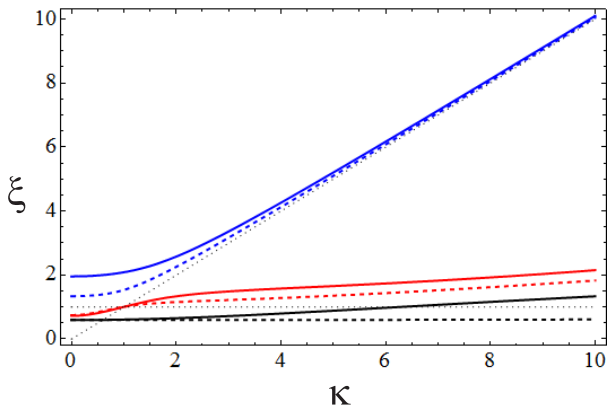}
\caption{\label{susdSOI 01} (Color online) Figure shows the spectrum of extraordinary waves in quantum plasmas. Continues lines describe three extraordinary waves existing in quantum plasma at the SSE account. The third extraordinary wave is the extraordinary SEAW presented by the lower (black) line. It exists due to the breaking of the symmetry between spin-up and spin-down states caused by the external magnetic field. The upper (blue) and middle (red) dashed lines describe the extraordinary waves with no account of the SSE and spin-orbit interaction. The lower (black) dashed line  describes the electrostatic SEAW existing due to consideration of the SSE. The horizontal dotted line shows $\xi=1$ which corresponds to $\omega=\omega_{Le}$. The inclined dotted line shows $\xi=\kappa$ which corresponds to $\omega=kc$. This figure is plotted for $n_{0e}=2.6\times10^{27}$ cm$^{-3}$, $B_{0}=10^{11}$ G, $\eta_{e}=0.09$, and temperature $T\ll T_{Fe}=5.9\times10^{7}$ K.}
\end{figure}
\begin{figure}
\includegraphics[width=8cm,angle=0]{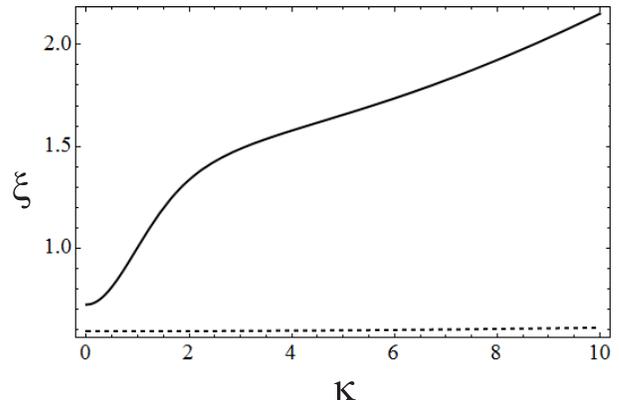}
\includegraphics[width=8cm,angle=0]{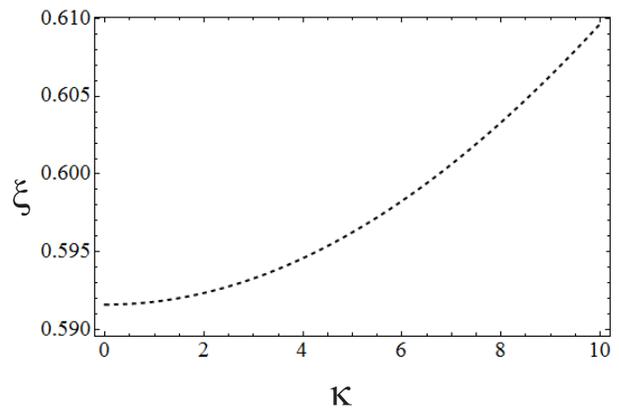}
\caption{\label{susdSOI 01 a} The upper figure shows the extraordinary SEAW (continuous line) and the electrostatic SEAW (dashed line) corresponding to the parameter regime of Fig. \ref{susdSOI 01}. The lower figure presents details of the dispersion dependence of the electrostatic SEAW presented in upper figure.}
\end{figure}
\begin{figure}
\includegraphics[width=8cm,angle=0]{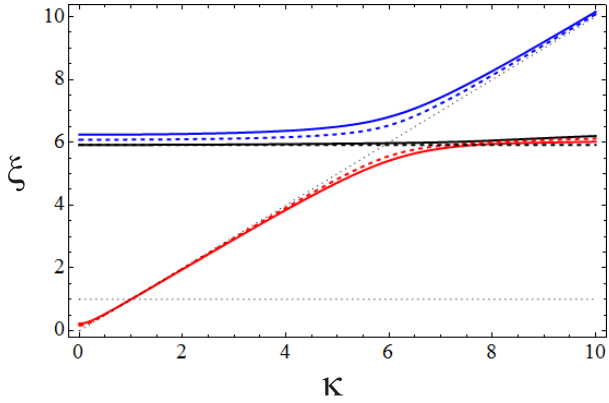}
\caption{\label{susdSOI 02} (Color online) Figure shows the spectrum of extraordinary waves in quantum plasmas. Description of the lines in this figure is similar to the description of lines in Fig. \ref{susdSOI 01}. However, the extraordinary SEAW is presented by the middle (black) continuous line in this picture. The figure is obtained for larger magnetic field $B_{0}=10^{12}$ G in compare with Fig. \ref{susdSOI 01}. It gives different spin polarization $\eta=0.72$. Concentration value is the same $n_{0e}=2.6\times10^{27}$ cm$^{-3}$.}
\end{figure}
\begin{figure}
\includegraphics[width=8cm,angle=0]{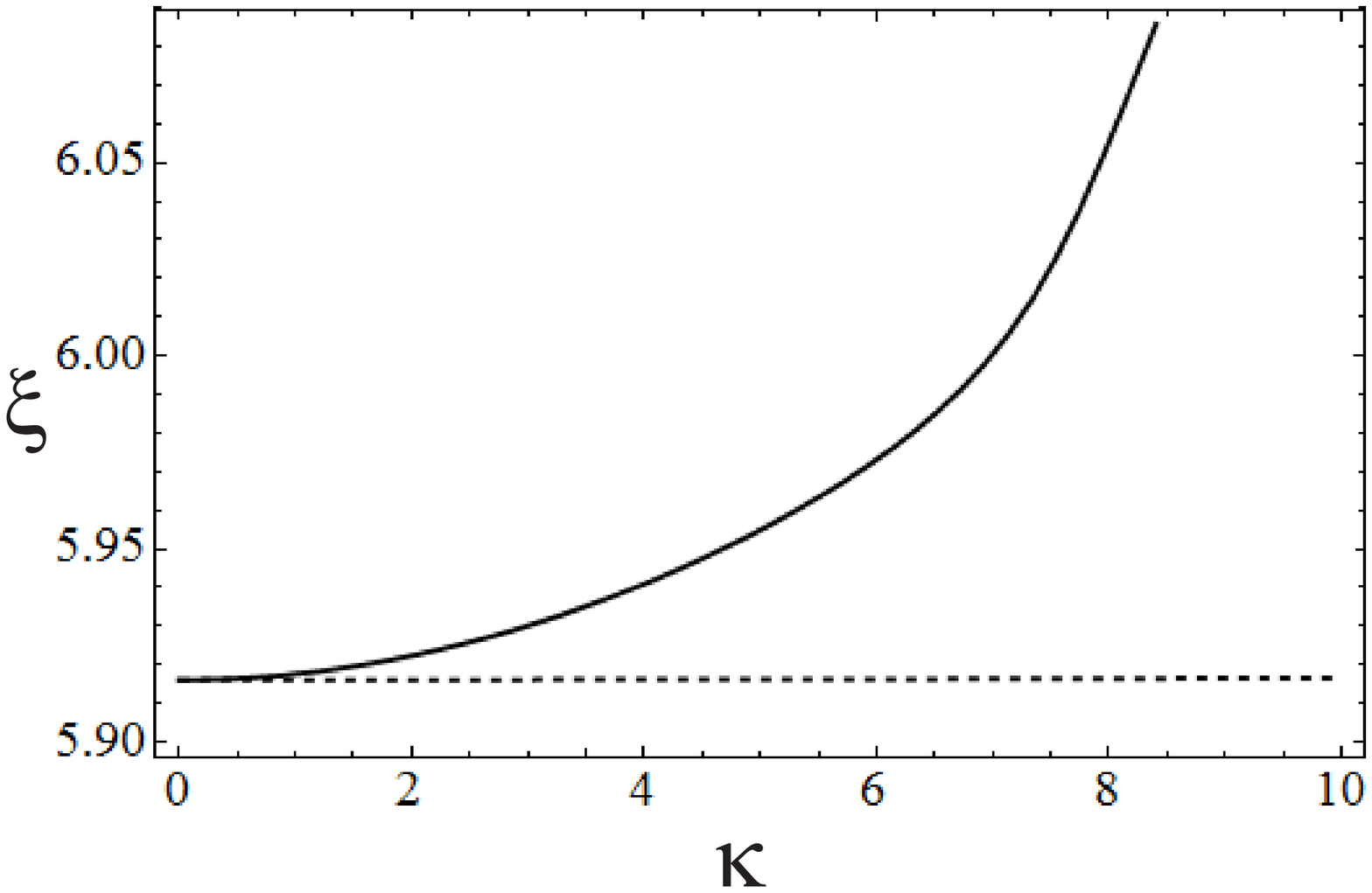}
\includegraphics[width=8cm,angle=0]{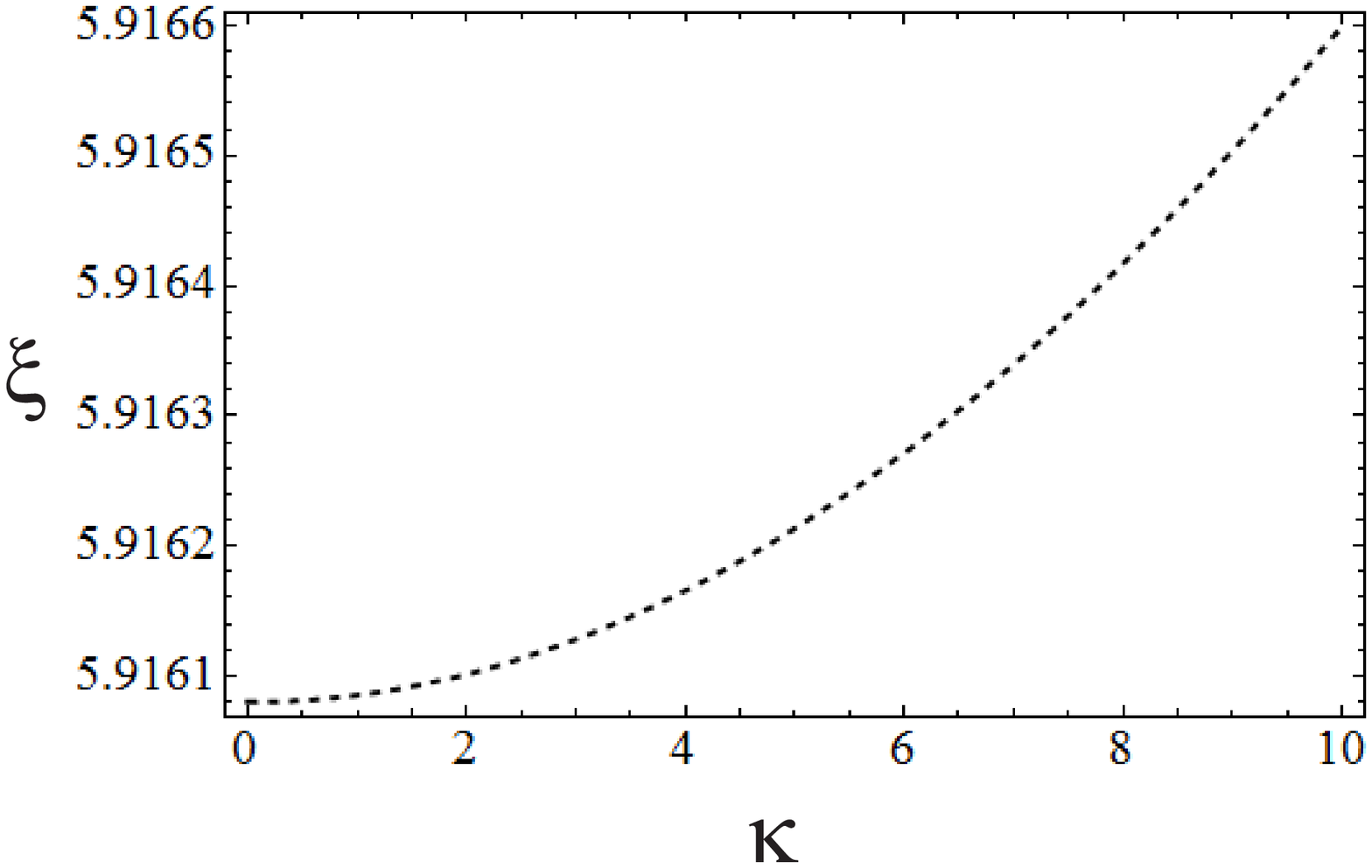}
\caption{\label{susdSOI 03} The upper figure shows the details of the dispersion dependence of extraordinary SEAW depicted in Fig. \ref{susdSOI 02}. The lower picture shows the electrostatic SEAW corresponding to the parameters used in the upper figure and Fig. \ref{susdSOI 02}.}
\end{figure}
\begin{figure}
\includegraphics[width=8cm,angle=0]{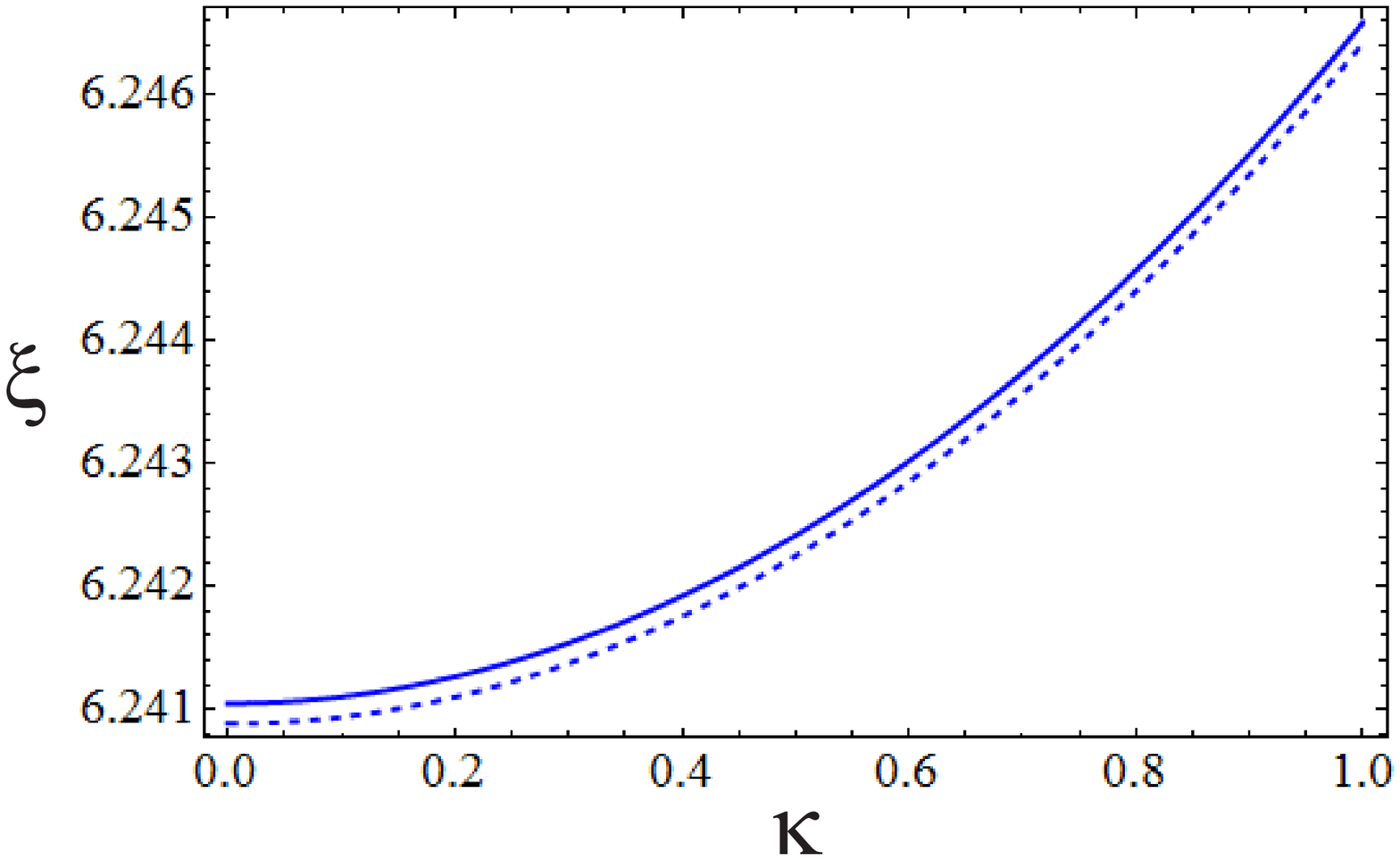}
\caption{\label{susdSOI 04} (Color online) The figure shows the upper extraordinary wave. The dashed line presents wave under influence of the SSE. The continuous line includes the effect of the spin-orbit interaction in addition to the SSE. Parameters are as in Fig. \ref{susdSOI 02}.}
\end{figure}
\begin{figure}
\includegraphics[width=8cm,angle=0]{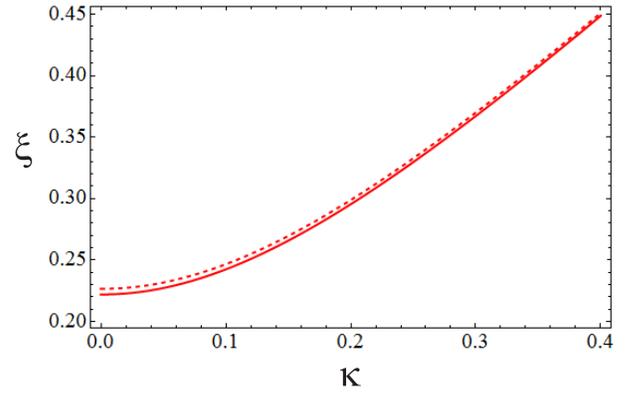}
\caption{\label{susdSOI 05} (Color online) The figure shows the lower extraordinary wave. The dashed line presents wave under influence of the SSE. The continuous line includes the effect of the spin-orbit interaction in addition to the SSE. Parameters are as in Fig. \ref{susdSOI 02}.}
\end{figure}

\subsection{Numerical analysis}

At the numerical analysis and presentation of our results we use the following dimensionless parameters $\xi=\omega/\omega_{Le}$, $\kappa=kc/\omega_{Le}$ and the following expression for the equilibrium spin polarization $\eta=\tanh(\mu_{B}B_{0}/\varepsilon_{Fe})$, where $\varepsilon_{Fe}=T_{Fe}/k_{B}=(3\pi^{2}n_{0e})^{\frac{2}{3}}\hbar^{2}/2m$ is the Fermi energy, and $k_{B}$ is the Boltzmann constant.

Usually, the dispersion equation for longitudinally-transverse waves propagating perpendicular to external field and having electric field in the plane perpendicular to the external magnetic field gives two solutions called the extraordinary waves. However, dispersion equation (\ref{susdSOI Disp eq}) has three solutions due to the SSE. The dispersion equation (\ref{susdSOI Disp eq}) also contains the spin-orbit interaction contribution.

The spin-orbit interaction is a semi-relativistic effect. To include its contribution we need to consider relatively large equilibrium concentrations $n_{0e}\sim 10^{27}$ cm$^{-3}$. As it was demonstrated and discussed earlier the exchange interaction plays considerable role at concentrations below $\approx 10^{25}$ cm$^{-3}$ \cite{Andreev PoP 16 soliton}, \cite{Andreev PP 15 Cylinder}.

Appearance of the third solution is similar to appearance of the longitudinal SEAWs \cite{Andreev PRE 15 SEAW}, \cite{Andreev AoP 15 SEAW}. Therefore, we call the third solution as the extraordinary SEAW.

Next, we need to study the dispersion dependence of three extraordinary waves and the influence of the spin-orbit interaction on their dispersion dependencies. Therefore, we consider the degenerate quantum plasma with $n_{0e}=2.6\times10^{27}$ cm$^{-3}$ and $B_{0}=10^{11}$ G. In this regime, the cyclotron frequency is smaller than the Langmuir frequency. Corresponding spectrum is presented in Fig. \ref{susdSOI 01}. Comparing upper (blue) continuous line and upper (blue) dashed line, we see that the account of the SSE leads to considerable increase of the frequency of the upper extraordinary wave. It includes an increase of the cut-off frequency at $k=0$. This shift becomes relatively small at the large wave vectors. The lower extraordinary wave shows different behavior. The SSE does not change its behavior at small wave vectors. Starting from $\kappa=1$ corresponding to the cross of $\omega=\omega_{Le}$ and $\omega=kc$, the dispersion of lower extraordinary wave considerably increases due to the SSE, as we see from comparison of the middle (red) continuous line and the middle (red) dashed line.

More detailed comparison of spectrums of the extraordinary SEAW and electrostatic SEAW is presented in upper figure in Fig. \ref{susdSOI 01 a}. Spectrum of the electrostatic SEAW is located above the electron cyclotron frequency. It is almost horizontal in the scale of upper figure in Fig. \ref{susdSOI 01 a}. While the spectrum of extraordinary SEAW is located in the area of considerably larger frequencies. It increases faster in compare with the electrostatic SEAW. Details of the spectrum of electrostatic SEAW are shown in lower figure in Fig. \ref{susdSOI 01 a}. We see that forms of dispersion dependencies of the extraordinary SEAW and electrostatic SEAW are also different.

At the next step, we consider the degenerate quantum plasma with $n_{0e}=2.6\times10^{27}$ cm$^{-3}$ and $B_{0}=10^{12}$ G. In this regime, the cyclotron frequency is larger than the Langmuir frequency. Corresponding spectrum is presented in Fig. \ref{susdSOI 02}. In this regime, the extraordinary SEAW is presented by the middle (black) continuous line. Similarly to the previously considered regime, we find that the SSE increases the frequency of upper extraordinary wave. We see it from the comparison of the upper (blue) continuous line and the upper (blue) dashed line. In opposite to the previous cases, the frequency of lower extraordinary wave decreases due to the SSE. As we see it from the comparison of the lower (red) continuous line and the lower (red) dashed line.

On this scale the extraordinary SEAW is almost horizontal line. To demonstrate details of its spectrum we plot the upper figure in Fig. \ref{susdSOI 03}. For comparison with the electrostatic SEAW we present the spectrum of the electrostatic SEAW in the lower figure in Fig. \ref{susdSOI 03}.

To consider the influence of the spin-orbit interaction on the extraordinary waves we present Figs. \ref{susdSOI 04} and \ref{susdSOI 05}.
The increase of frequency of the upper extraordinary wave and the decrease of frequency of the lower extraordinary wave at relatively small wave vectors are demonstrated in Figs. \ref{susdSOI 04} and \ref{susdSOI 05}, correspondingly. It decreases at the larger wave vectors.

The contribution of spin-orbit interaction in the chosen parameter regime
$B_{0}=10^{11}$ G and $B_{0}=10^{12}$ G $\ll B_{cr}=m_{e}^{2}c^{3}/(e\hbar)=4.41\times10^{13}$ G corresponding to $\hbar\Omega\leq mc^{2}$ and
$\varepsilon_{Fe}=0.01 mc^{2}$ is relatively small, so we cannot show in general
pictures (Fig. \ref{susdSOI 01} and \ref{susdSOI 02}). At the larger concentrations,
the contribution of spin-orbit interaction would be larger. However, we work in
regime of parameters corresponding to the semi-relativistic approach in accordance
with the area of applicability of our equations.

\section{Conclusions}

Generalization of the NLPE containing the spinor pressure contribution has been constructed to include the effect of spin-orbit interaction. Corresponding generalization of the SSE-QHD containing the pair of continuity equations, pair of Euler equations for spin-up and spin-down electrons and equations of the spin evolution for $S_{x}$ and $S_{y}$ projections of the spin density has been constructed either.

The spin-orbit interaction is an interesting example of quantum-relativistic effects. Its analysis presented in this paper is an important step towards the construction of quantum-relativistic statistical model in which the temperature is a local characteristic of medium \cite{MaksimovTMP 2001} in contrast with the external parameter quantum statistics based on the Gibbs measure. Presented here model is a limit case of many-particle QHD located on the SSE in the degenerate electron gas.

Application of developed here model to the wave phenomena in the electron gas show three results.

First, we have shown the contribution of the SSE in the spectrums of extraordinary waves. Second, a modification of the spectrum of SEAWs propagating perpendicular to the external magnetic field approximately described earlier as a longitudinal wave is obtained at the account of the transverse electric field contribution. We have found considerable change of spectrum of all three extraordinary waves. Third, we have demonstrated extra shifts of the dispersion dependencies of extraordinary waves under the influence of the spin-orbit interaction.

\section{Appendix A: Spin current}

We need to have the explicit form of the spin-current tensor components to recognize their parts in the terms caused by the spin-orbit interaction. Therefore, we present their nonrelativistic parts here:
\begin{equation}\label{susdSOI} J^{x\alpha}=\frac{1}{2m}(\Phi_{u}^{*}D_{\alpha}\Phi_{d} +\Phi^{*}_{d}D_{\alpha}\Phi_{u}+c.c.), \end{equation}
\begin{equation}\label{susdSOI}J^{y\alpha}= \frac{1}{2m}(\Phi^{*}_{u}(-\imath)D_{\alpha}\Phi_{d} +\Phi^{*}_{d}\imath D_{\alpha}\Phi_{u}+c.c.), \end{equation}
\begin{equation}\label{susdSOI}J^{z\alpha}= \frac{1}{2m}(\Phi^{*}_{u}D_{\alpha}\Phi_{u} -\Phi^{*}_{d} D_{\alpha}\Phi_{d}+c.c.). \end{equation}

\section{Appendix B: Equation of state for $j^{\alpha\beta\gamma}$}
An explicit expression for $j^{\alpha\beta\gamma}$ defined by formula (\ref{susdSOI def j}) is equal to the explicit expression for the spin current flux $J^{\alpha\beta\gamma}$ defined by formula (\ref{susdSOI def J}) up to the quantum terms similar to the quantum Bohm potential.
Qualitatively speaking, the spin current flux is the average value of the following quantity: $J^{\alpha\beta\gamma}=\langle S^{\alpha}_{i}v^{\beta}_{i}v^{\gamma}_{i}\rangle$, where $S^{\alpha}_{i}$ and $v^{\alpha}_{i}$ are the spin and velocity of i-th particle. Splitting spin and velocity on the value related to the local center of mass and the value related to the thermal motion $S^{\alpha}_{i}=S^{\alpha}(\textbf{r},t)+s^{\alpha}_{i}$ and $v^{\alpha}_{i}=v^{\alpha}(\textbf{r},t)+u^{\alpha}_{i}$ we find $J^{\alpha\beta\gamma}=S^{\alpha}(\textbf{r},t)v^{\beta}(\textbf{r},t)v^{\gamma}(\textbf{r},t)+S^{\alpha}(\textbf{r},t)\langle u^{\beta}_{i}u^{\gamma}_{i}\rangle +v^{\beta}(\textbf{r},t)\langle s^{\alpha}_{i}u^{\gamma}_{i}\rangle
+v^{\gamma}(\textbf{r},t)\langle s^{\alpha}_{i}u^{\beta}_{i}\rangle +\langle s^{\alpha}_{i}u^{\beta}_{i}u^{\gamma}_{i}\rangle$ $=S^{\alpha}(\textbf{r},t)v^{\beta}(\textbf{r},t)v^{\gamma}(\textbf{r},t)+S^{\alpha}(\textbf{r},t) p^{\beta\gamma}+v^{\beta}(\textbf{r},t)J^{\alpha\gamma}_{th}
+v^{\gamma}(\textbf{r},t)J^{\alpha\beta}_{th} +J^{\alpha\beta\gamma}_{th}$, where $J^{\alpha\beta}_{th}$ and $J^{\alpha\beta\gamma}_{th}$ are the thermal spin current and the thermal spin current flux.

For our calculations we need equilibrium value of $j^{\alpha\beta\gamma}$. It means we need the equilibrium spin current flux $J_{0}^{\alpha\beta\gamma}=0+\delta^{z\alpha}S^{z}(\textbf{r},t) p_{Fe}\delta^{\beta\gamma} +0+0+J^{\alpha\beta\gamma}_{0,th}$. For the calculation of the equilibrium thermal spin current flux $J^{\alpha\beta\gamma}_{0,th}$ we can use the equilibrium spin distribution functions obtained in Ref. \cite{Andreev 1409} for SSE quantum kinetics $J^{\alpha\beta\gamma}=\int p^{\gamma}p^{\beta}S^{\alpha}(\textbf{r},\textbf{p},t)d\textbf{p}=\delta^{z\alpha}S^{z}(\textbf{r},t) p^{\beta\gamma}$. It is already included in $J_{0}^{\alpha\beta\gamma}$. It means $J^{\alpha\beta\gamma}_{0,th}=0$. Consequently, we find $j^{\alpha\beta\gamma}_{0}=\delta^{z\alpha}\delta^{\beta\gamma}S^{z} p_{Fe}$ which gives zero contribution in the linear evolution (\ref{susdSOI Euler lin u}).

\begin{acknowledgements}
The authors thank Professor L. S. Kuz'menkov for fruitful discussions. The work of P.A. was supported by the Russian
Foundation for Basic Research (grant no. 16-32-00886) and the Dynasty foundation.
\end{acknowledgements}

\end{document}